\documentclass{PoS}
\usepackage{subfig}
\usepackage{graphicx}
\usepackage{xcolor}
\usepackage{slashed}\usepackage{slashed}
\definecolor{darkgreen}{rgb}{0.0,0.6,0.0}
\definecolor{cDarkGrey}{RGB}{91,91,91}
\definecolor{cGrey}{RGB}{245,243,238}
\definecolor{cBlue}{RGB}{0,110,191}
\definecolor{cLightBlue}{RGB}{214,237,252}
\definecolor{cRed}{RGB}{196,0,100}
\definecolor{cLightRed}{RGB}{254,222,237}
\definecolor{cGreen}{RGB}{0,166,80}
\definecolor{cLightGreen}{RGB}{254,222,237}
\definecolor{cOrange}{RGB}{221,74,44}
\definecolor{cLightOrange}{RGB}{255,215,210}
\definecolor{cPurple}{RGB}{93,35,125}
\definecolor{cLightPurple}{RGB}{241,230,252}
\definecolor{cYellow}{RGB}{252,191,10}
\definecolor{cISSRBlue}{RGB}{0,111,174}
\definecolor{cISSRGrey}{RGB}{167,169,172}
\title{Next-to-leading power threshold factorization for  Drell-Yan production}

\ShortTitle{NLP threshold factorization for DY production}

\author{ \speaker{Sebastian Jaskiewicz}\thanks{Preprint number: TUM-HEP-1246/19}\\
	Physik Department T31,
	Technische Universit\"at M\"unchen,\\
	James-Franck-Stra{\ss}e 1,  D-85748 Garching, Germany\\
	E-mail: \email{sebastian.jaskiewicz@tum.de}}

\abstract{We present the next-to-leading power (NLP)  factorization formula for the $q\bar{q}\to \gamma^*+X$ channel of  the Drell-Yan 
	production near the kinematic threshold limit. The formalism used for the computation of next-to-leading power corrections within soft-collinear effective field theory  
	is introduced, we discuss the emergence of new objects, the {\it{NLP collinear functions}}, and define them through an operator matching equation.  
 We review the leading power factorization before extending it 
	to subleading powers. We also present the one-loop result for the newly introduced collinear function, 
	and demonstrate explicitly conceptual issues in performing next-to-leading 
	logarithmic resummation at next-to-leading power. }

\FullConference{14th International Symposium on Radiative Corrections (RADCOR2019)\\ 
9-13 September 2019\\
		Palais des Papes, Avignon, France\\
		}

\begin{document}

\section{Introduction}
A detailed description of the soft-collinear factorization for the Drell-Yan (DY) 
process in the kinematic threshold limit at subleading powers
has recently been  presented in   
\cite{Beneke:2019oqx}. With this in mind, the aim of this
contribution is to focus on the key ideas and results while
keeping the technical details and subtleties to a minimum as much as it is possible.
For these, we encourage the interested reader to consult \cite{Beneke:2019oqx} and \cite{Beneke:2018gvs}.

The partonic process we consider is  $q\bar q\to\gamma^*(Q)+X$ at threshold,
that is the $z=Q^2/\hat{s}\to 1$ limit,  where $\hat{s}=x_ax_b \,s$ is the partonic 
centre-of-mass energy  squared, $x_a,x_b$ are the momentum 
fractions of the partons inside the incoming hadrons and $Q^2$ is 
the invariant mass squared of the off-shell photon. The final state QCD radiation $X$ is forced to be soft and
the cross-section is expanded in  powers of $(1-z)$.  
The main results of this work, \cite{Beneke:2019oqx}, and \cite{Beneke:2018gvs} are first, the derivation of the 
factorization formula at next-to-leading power (NLP) in the threshold expansion, and second,
the identification of new physical objects that emerge beyond leading power:
 the amplitude level {\it{NLP collinear functions}} 
and, in addition, the generalized soft functions previously defined in \cite{Beneke:2004in}.


Recently, interest in subleading-power
 corrections has arisen in the theoretical community.
Subleading-power effects are important to consider
both from a theoretical and a phenomenological perspective, in order
to advance the understanding of all-order structure of quantum field theories, and improve subtraction methods. 
Computations of subleading-power corrections contribute
to providing precise predictions for processes within
the Standard Model (SM) and they can be numerically 
important.  This was recently shown in the 
 study of the leading logarithmic NLP corrections in 
Higgs prouction via gluon fusion \cite{Beneke:2019mua}. 
These considerations are crucial in matching
the experimental efforts, in particular in the upcoming
era of HL-LHC. 
The conceptual leap to next-to-leading 
logarithmic accuracy at NLP is key as 
 current efforts to extend the subleading power 
resummation beyond the leading logarithmic order 
have been hampered by the issue of endpoint divergences in 
ill-defined convolutions 
\cite{Beneke:2019oqx,Beneke:2017vpq,Beneke:2019slt,Alte:2018nbn,Moult:2019uhz}. 
We make this issue explicit  by considering the results of 
the computations presented here. 
\section{SCET formalism}
\label{sec:SCET}
In this work, the position-space formulation
\cite{Beneke:2002ph,Beneke:2002ni} of soft-collinear effective theory (SCET) 
\cite{Bauer:2000yr,Bauer:2001yt,Bauer:2001ct} is used. The SCET Lagrangian is split 
into $N$ collinear sectors, denoted by a subscript $c_i$, which interact with each other only through the 
exchange of soft partons. Namely, it is given by
\begin{equation}
\mathcal{L}_{{\rm{SCET}}}= \sum^N_{i=1}\,\mathcal{L}^{}_{c_i} +\mathcal{L}^{}_{{\rm{soft}}}\,, 
\end{equation}
where each of the Lagrangians belonging to a collinear direction is expanded in powers of the small power-counting parameter $\lambda$. In the present case the small power-counting parameter is $\lambda=\sqrt{1-z}$ which corresponds to the threshold-collinear scale: 
 \begin{equation}
		\,\mathcal{L}^{}_{c_i}= \underbrace{ \mathcal{L}^{(0)}_{c_i}}_{\rm{LP}} + \underbrace{\mathcal{L}^{(1)}_{c_i}}_{\mathcal{O}(\lambda^1)} + \underbrace{\mathcal{L}^{(2)}_{c_i}}_{\mathcal{O}(\lambda^2)}+\,\,... \nonumber  
		\end{equation}  
The first term in this expansion constitutes the leading power (LP) contribution, and the terms which follow are the subleading-power corrections. 
This formalism can in general describe $N$-jet processes. In the case of the DY production we set $N=2$, resulting in  collinear, $c$, and anticollinear, $\bar{c}$, sectors.
The LP collinear Lagrangian is given by   \cite{Beneke:2002ni}
\begin{equation}
	\label{eq:LagrangianLP}
		{\mathcal{L} }^{(0)}_{c} = \bar{\xi}_c \left(i n_- D_c+ g_s\, \textcolor{red}{n_-  A_s (z_-)} + i \slashed{D}_{\perp c}
		\frac{1}{i n_+ D_{c}}\, i\slashed{D}_{\perp c} \right)
		\frac{\slashed{n}_+}{2} \, \xi_c + \mathcal{L}^{(0)}_{c,{\rm{YM}}}
		\,,
\end{equation}
with  $i n_- D_c = i n_-  \partial + g_s\, \textcolor{blue}{n_-  A_c (z)} $. $\mathcal{L}^{(0)}_{c,{\rm{YM}}}$ is the LP soft-collinear Yang-Mills Lagrangian, the unspecified arguments of collinear fields are the full $z$ coordinates, $n^{\mu}_-, n^{\mu}_+$ are light-like vectors with $n_-\cdot n_+=2$, and $\textcolor{red}{A_s}$ is evaluated at position $z_-^{\mu}= (n_+z){n_-^{\mu}}/{2}$  as a consequence of multipole expansion.
The soft-collinear interaction at LP is given by the standard eikonal vertex.     		

Importantly, at LP, we can apply the decoupling 
transformation \cite{Bauer:2001yt}, for example for the initial state collinear quark:\footnote{For the gauge field the decoupling transformation is given by $A^\mu_c(z) \to  Y_+(z_-) A^{(0) \mu}_{c}(z)Y^\dagger_+(z_-)$.} 
$\xi_c(z) \to Y_+(z_-) \, \xi_c^{(0)}(z)$, 
where 
\begin{eqnarray}
Y_{\pm}\left(x\right)&=&\mathbf{P}
\exp\left[ig_s\int_{-\infty}^{0}ds\,n_{\mp}
A_{s}\left(x+sn_{\mp}\right)\right],
\end{eqnarray}
is the soft Wilson line. This field redefinition 
removes all the interactions between the soft and collinear fields from the LP Lagrangian in (\ref{eq:LagrangianLP})
as  
\begin{eqnarray}
\bar{\xi}_c \,(i n_- D_c + g_s\textcolor{red}{n_-  A_s (x_-)})\, \frac{\slashed{n}_{+}}{2}\,\xi_c 
= \bar{\xi}_c^{(0)} \, in_-D_c^{(0)}\, \frac{\slashed{n}_{+}}{2}\,
\xi_c^{(0)}\,,
\end{eqnarray}
where the superscript $(0)$ denotes the decoupled fields. 
	
We next consider the SCET formalism beyond the leading power. 
The systematic expansion  in powers of $\lambda$ which is built into the 
SCET framework means that this effective field theory is ideally suited to the study
of power corrections. The formalism we use here was developed in 
\cite{Beneke:2004in,Beneke:2017ztn,Beneke:2018rbh,Beneke:2019kgv}.\footnote{See \cite{Moult:2019uhz,Feige:2017zci,Chang:2017atu,Ebert:2018lzn,Moult:2017rpl,Moult:2018jjd} for alternative approach using label formalism.} 
A generic, $N$-jet, operator has the following form 
\begin{eqnarray}
	J = \int dt \,C(\{t_{i_k}\})J_s(0)\prod_{i=1}^{N}J_{i}(t_{i_1},t_{i_2},...).
\end{eqnarray}
It is a product of $N$ operators, $J_i$, each of which associated with 
a particular collinear direction, and a soft operator, $J_s$, which consists only of soft fields. 
The measure is given by $dt=\prod _{{ik}}dt_{i_k}$ and $C(\{t_{i_k}\})$ is a hard matching coefficient.
Each of the $J_i$ operators is constructed from 
collinear-gauge-invariant collinear building blocks  		
\begin{eqnarray}
	J_i(t_{i_1},t_{i_2},...) = \prod_{k=1}^{n_i}\psi_{i_k}(t_{i_k}n_{i+})\,,
\end{eqnarray}		
which are given by 
	\begin{equation}
	\label{eq:Vccsc}	\psi_{i}(t_in_{i+}) \in  
	\,\left\{ \begin{array}{ll}
	\displaystyle  \chi_i(t_in_{i+})
	\equiv W_i^{\dagger}
	\,\xi_i  & \quad{\rm{collinear\,quark} } 
	\\[0.3cm]
	\displaystyle  \mathcal{A}^{\mu}_{i\perp}(t_in_{i+})
	\equiv W_i^{\dagger}
	\left[i\,D^{\mu}_{i\perp}\,W_i \right]
	& \quad{\rm{collinear\,gluon} }	\end{array}\right.
	\end{equation}
Each of the building blocks scales as $\mathcal{O}(\lambda)$. In the above equation, we make use of the i$^{\rm{th}}$-collinear 
Wilson line, which is a path-ordered exponential 
of $n_{i+}A_{i}$. For DY, we require $i=c$  which is given by
\begin{eqnarray} 
W_{c}\left(x\right)&=&\mathbf{P}
\exp\left[ig_s\int_{-\infty}^{0}ds\,n_+
A_{c}\left(x+sn_+\right)\right],
\end{eqnarray} 
and a corresponding definition for the $\bar{c}$-direction
 with $n_+\leftrightarrow n_-$. 	
	
The LP configuration is simply given by the presence of one building
block in each of the $N$ collinear directions, 
\begin{eqnarray}
	J^{A0}_i(t_i)=\psi_{i}(t_in_{i+}).
\end{eqnarray}
There is a number of ways to include power suppression in this formalism. The first, is to introduce derivatives, $\partial^{\mu}_{\perp}\sim\lambda$,  that act on the fields present in the LP configuration. The second way, is to place additional building blocks in a particular collinear direction, each building block gives a power of $\lambda$ suppression. This procedure gives rise to subleading power currents such as $\textcolor{cRed}{J^{An}_i},\textcolor{cBlue}{J^{Bn}_i},\textcolor{cGreen}{J^{Cn}_i}$. They are labelled as follows: 
\begin{itemize}
	\item $\textcolor{cRed}{A},\textcolor{cBlue}{B},\textcolor{cGreen}{C}...$ refers to number of fields in a given collinear direction
	\item $n$ is the power of $\lambda$ suppression (relative to $A0$) in a given sector.
\end{itemize}
Examples of such currents are 
\begin{eqnarray}
\nonumber \textcolor{cRed}{i\partial^{\mu}_{\perp i}i\partial^{\nu}_{\perp i}\chi_i}, \hspace{0.4cm} \textcolor{cBlue}{\chi_i(t_{i_1})i\partial^{\nu}_{\perp i}\mathcal{A}^{\mu}_{i\perp}(t_{i_2})}, \hspace{0.4cm} \textcolor{cGreen}{\chi_i(t_{i_1})\chi_i(t_{i_2})\mathcal{A}^{\mu}_{i\perp}(t_{i_3})},\hspace{0.4cm} 
\end{eqnarray}
for $\textcolor{cRed}{J^{A2}_i}, \textcolor{cBlue}{J^{B2}_i},$ and $\textcolor{cGreen}{J^{C2}_i}$ respectively.
The overall power suppression for the $N$-jet operator is given by the sum of the power suppression from the different sectors. Hence, an $\mathcal{O}(\lambda^2)$ 2-jet operator could be given by the following
\begin{eqnarray}
\nonumber J^{A2}_1J^{A0}_2, \hspace{1cm}J^{A1}_1J^{B1}_2,\hspace{1cm}J^{A0}_1J^{C2}_2,\,\, ... 
\end{eqnarray}
In addition to the currents described above, there is also the possibility to form time-ordered products of subleading-power Lagrangian terms with lower power currents in order to provide power suppression. These contributions, $\textcolor{cOrange}{J^{Tn}_i}$, are labelled by the letter $T$ and  a number $n$ which gives the total power suppression from the current and the insertions of the subleading-power Lagrangian terms. An example is
\begin{eqnarray}
 \textcolor{cOrange}{J^{T2}_i(t_{i_1})=i\int d^4z \,\mathbf{T}\big[\chi_i(t_{i_1})\mathcal{L}^{(2)}_i(z)\big]},
\end{eqnarray}
where $\mathcal{L}^{(2)}_i(z)$ is a $\lambda^2$ power-suppressed Lagrangian term \cite{Beneke:2002ni}.
These time-ordered product insertions play a crucial role the derivation of the factorization formula for the threshold DY  production at NLP, and we will investigate them in much greater detail in the later sections. 
\section{Factorization at leading power}
Having briefly introduced the subleading-power SCET formalism
we turn our attention to the specific case of the DY process at 
threshold. We will first review the LP factorization, and later
investigate the subleading-power effects. The partonic DY proccess at threshold 
is described by a standard SCET$_{\rm{I}}$ set up. In terms of the power-counting 
parameter $\lambda$, defined above, the threshold-collinear modes 
are given by   
\begin{eqnarray}
\textcolor{blue}{p_c} & =& (n_+p_c,n_-p_c,p_{c\perp}) \sim Q(1,\lambda^2,\lambda), \nonumber \\
\textcolor{darkgreen}{p_{\bar{c}}} &=& (n_+p_{\bar{c}},n_-p_{\bar{c}},p_{{\bar{c}}\perp}) \sim Q(\lambda^2,1,\lambda), \nonumber  \\
\textcolor{red}{p_s} &=& (n_+p_s,n_-p_s,p_{s\perp}) \sim Q(\lambda^2,\lambda^2,\lambda^2) .
\end{eqnarray}  
In addition to these modes, at hadronic level there exist the (anti)collinear-PDF 
modes, with the transverse momentum scaling of $p_{\perp}\sim\Lambda$, where
$\Lambda$ is the scale of strong interactions. The $c$-PDF modes have momentum scaling  
$(Q,\Lambda^2/Q,\Lambda)$. The usual parton distribution functions
(PDFs) are described by these modes. Here we consider the power corrections in $\lambda$ and work at leading power in $\Lambda$. The set up is perturbative as the 
threshold-collinear scale is still much larger than the scale of strong interactions $\Lambda$, 
$Q\lambda=Q\sqrt{(1-z)} \gg  \Lambda$. 

\begin{figure}[t]%
	\centering
	\subfloat[Amplitude representation prior to use of decoupling transformation.]{{\includegraphics[width=4cm]{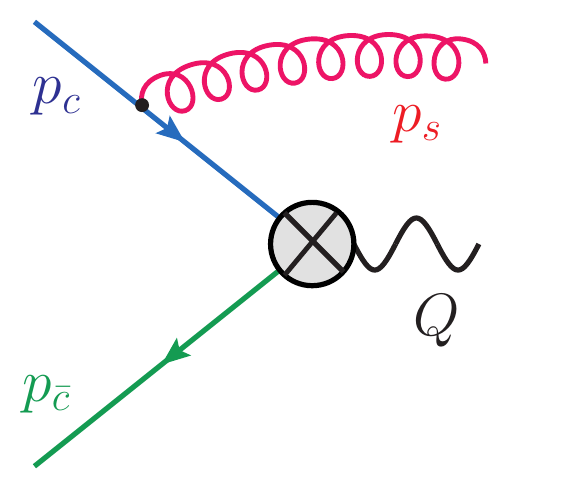} }}%
	\qquad
	\subfloat[Schematic representation {\it{after}} decoupling transformation is performed.]{{\includegraphics[width=4cm]{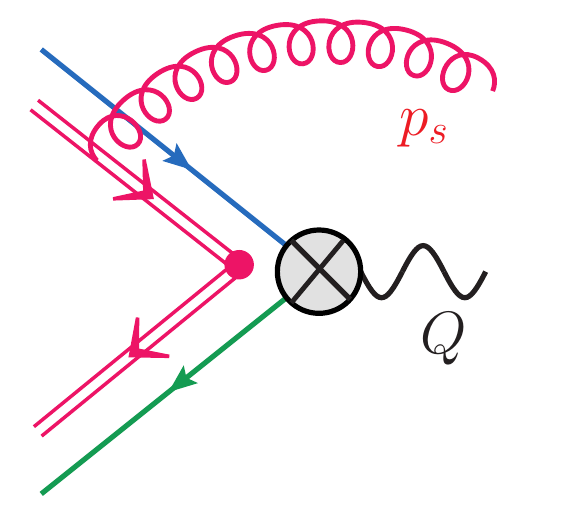} }}%
	\caption{Representation of the LP amplitude before and after the decoupling transformation.
		}%
	\label{fig:example}%
\end{figure}

The first step in the derivation of factorization of DY  in SCET  at LP 
\cite{Becher:2007ty} involves matching the electromagnetic quark current to the LP SCET current,  
\begin{eqnarray}\label{eq:LPmatching}
\bar{\psi} \gamma_\rho \psi(0) = \int dt\, 
d\bar{t}\,\widetilde{C}^{A0,A0}(t,\bar{t}\,) \,
J_\rho^{A0,A0}(t,\bar{t}\,)
\end{eqnarray}
where in our notation ($A0$ labelling the LP currents in the collinear and anticollinear directions)
\begin{eqnarray}\label{eq:LPcurrent}
J_\rho^{A0,A0}(t,\bar{t}\,) =  \bar{\chi}_{\bar{c}} 
(\bar{t} n_-)\gamma_{\perp\rho} \chi_c(t n_+ )\,
\end{eqnarray}
before the decoupling transformation
\cite{Bauer:2001yt} is used. The calculation proceeds by taking the matrix element of the above operator for the incoming (anti)collinear (anti)quark and the final state QCD radiation, $X$. Performing now the decoupling transformation, the states factorize and one obtains  
\begin{eqnarray}
\label{eq:2.2}
\langle X|\bar{\psi}\, \gamma^{\,\rho} \psi(0)|A(p_A)B(p_B)\rangle
&=&
\int dt\, d\bar{t}\,  C^{A0,A0}(t, \bar{t}\,)\,
\langle X^{{\rm PDF}}_{\bar{c}}|\bar{\chi}_{\bar{c}}
\,(\bar{t} n_-)|B(p_B)\rangle  \,
\gamma_{\perp}^{\,\rho} \nonumber\\ 
&& \hspace*{-0cm}\times\, 
\langle X^{{\rm PDF}}_{{c}}|
\chi_{c}\left(tn_+\right)
|A(p_A)\rangle\,\langle X_s|   {\mathbf{ T}}
\left( \left[ Y_{-}^\dagger(0) Y_{+}(0) \right]_{}
\right)|0 \rangle\,.
\end{eqnarray}
A picture of the factorization at amplitude level is presented in Fig.~\ref{fig:example}.
We note here that the final state radiation can be soft, $X_s$ and PDF-collinear, $X^{{\rm PDF}}_{{c}}$, but not threshold-collinear. This is an important point, to which we will return to below. However, in the LP calculation the usual steps follow.
Upon squaring of the above matrix element, performing the sum over the final state radiation, and combining the hadronic result with the lepton tensor gives the following  
\begin{equation}
\frac{d\sigma_{\rm DY}}{dQ^2} = 
\frac{4\pi\alpha_{\rm em}^2}{3 N_c Q^4}
\sum_{a,b} \int_0^1 dx_a dx_b\,f_{a/A}(x_a)f_{b/B}(x_b)\,
\hat{\sigma}^{{\,\rm{LP}}}_{ab}(z)\,.
\label{eq:dsigsq2}
\end{equation}
The $f_{a/A}(f_{b/B})$ is the standard PDF formed by the square of the (anti)collinear matrix element in~(\ref{eq:2.2}). The focus of our investigations is the perturbative factorization of the LP partonic cross section. 
At LP, the partonic cross section $\hat{\sigma}^{{\,\rm{LP}}}_{ab}(z)$ factorizes 
into a hard function, originating from squaring the hard matching 
coefficient $C^{A0,A0}(t, \bar{t}\,)$  and a
soft function:
\begin{equation}
\hat{\sigma}^{{\,\rm{LP}}}(z) = H(Q^2) \,Q S_{\rm DY}(Q(1-z))\,.
\label{eq:LPfact}
\end{equation}
The leading power DY soft function is given by the vacuum matrix element of only the soft Wilson lines
\cite{Korchemsky:1993uz}
\begin{equation}
S_{\rm DY}(\Omega) = \int \frac{dx^0}{4\pi}\,e^{i \Omega\, x^0 /2}\,
\frac{1}{N_c}\,\mbox{Tr} \,
\langle 0|\mathbf{\bar{T}}(Y^\dagger_+(x^0) Y_-(x^0)) 
\,\textbf{T}(Y^\dagger_-(0) Y_+(0))
|0\rangle\,.
\label{eq:LPsoftfn}
\end{equation}
Before starting the discussion of extending this factorization formula to subleading powers, we would like to draw attention to the simplicity 
of the LP result in (\ref{eq:LPfact}). There is no collinear dynamics present and the result is simply a product of the hard function and one soft function. 
\section{Factorization at next-to-leading power}
Thus far we have reviewed the LP factorization 
which is already well established in the literature \cite{Moch:2005ky,Becher:2007ty}. 
In this section we take the step beyond, which is 
to consider the factorization of the partonic 
cross section at NLP. We denote this contribution by 
$\hat{\sigma}^{{\rm{NLP}}}_{ab}(z)$
and it replaces $\hat{\sigma}^{{\rm{LP}}}_{ab}(z)$
in Eq.~(\ref{eq:dsigsq2}). We will first present 
the result in a schematic way since we wish to highlight the new features 
that are present. In the following sections we will motivate
their appearance and write down their precise form. 

We begin by stating that the NLP partonic cross 
section is given by 
\begin{equation}
	\hat{\sigma}^{{\,\rm{NLP}}}  = \sum_{\rm{terms}}\, [C\otimes\textcolor{blue}{J}\otimes \textcolor{darkgreen}{\bar{J}} \,\,]^2 
	\otimes \textcolor{red}{S}\,,
	\label{eq:factsketch}
	\end{equation}
where  $C$ is the hard Wilson matching coefficient, $\textcolor{red}{S}$ is a {\emph{generalized}} soft function
and  $\textcolor{blue}{J}$ is a {\it{NLP collinear function}}. The sum over ``terms'' means the sum over all the possible ways of inducing power suppression in the partonic cross section. Let us now motivate the emergence of this structure at next-to-leading power. 
\subsection{NLP collinear functions}
The first obvious difference in the schematic NLP cross section in (\ref{eq:factsketch}) with respect to the LP result in (\ref{eq:LPfact}) is the presence of the amplitude level {\it{NLP collinear functions}}, $\textcolor{blue}{J}$. 
These are indeed a new physical feature of the factorization starting at NLP. We would like to first explain why these collinear functions have not played a part in the derivation of the LP factorization formula and the reason for their appearance here. A detailed argument has been presented in \cite{Beneke:2019oqx} and here we outline the main ideas.
\begin{figure}[t]
	\begin{centering}
		\includegraphics[width=0.35\textwidth]{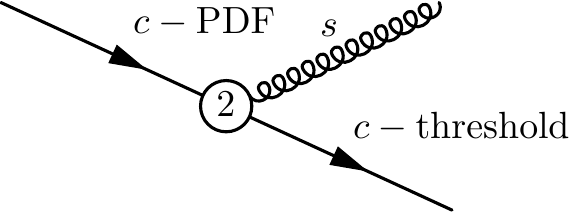}
		\par\end{centering}
	\caption{\label{fig:illustration}  
		Insertion of the power-suppressed Lagrangian 
		$\mathcal{L}^{(2)}_{2\xi}$ into a collinear quark line, which transforms a $c$-PDF quark into a threshold-collinear quark.}
\end{figure}
\begin{figure}[t]
	\begin{centering}
		\includegraphics[width=0.55\textwidth]{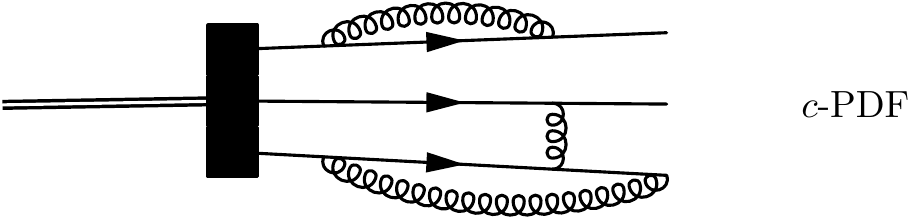}
		\par\end{centering}
	\caption{\label{fig:img2}  		
		The loops depicted in the figure have threshold-collinear 
		scaling. However, since at LP the matrix element does not have 
		support for these modes, they can be trivially integrated
		out, $\chi_c\to {\chi}^{{\rm PDF}}_{c }$. 	}
\end{figure}

The key observation at LP is that the decoupling transformation completely removes the soft-collinear interactions in the LP Lagrangian as was stressed in Sec.~\ref{sec:SCET}. Therefore, diagrams such as the one presented in Fig.~\ref{fig:illustration} do not feature in the calculations. This means that  in the on-shell matching to the $c$-PDF fields, loops formed only by the threshold-collinear fields can appear.  These are scaleless and therefore zero to all orders in perturbation theory in dimensional regularization. 
 This is represented in Fig.~\ref{fig:img2}. 
Hence, in  practice, the threshold-collinear modes are trivially integrated out and the threshold-collinear fields are simply identified with $c$-PDF fields, $\chi_c\to {\chi}^{{\rm PDF}}_{c }$. For this reason, we have not discussed 
 collinear functions in the derivation of the LP factorization formula between equations (\ref{eq:2.2}) and (\ref{eq:dsigsq2}). The collinear functions are in fact present there, but they are delta functions to all orders in perturbation theory: 
	\begin{eqnarray}
	\chi_c(tn_+) = \int\, du\,\tilde{J}(t,u)\, 
	\chi_c^{\rm{PDF}}(un_+)
	\end{eqnarray}
	with the collinear function written in position space $\tilde{J}(t,u) = \delta(t-u)$. Therefore, as stated,  the computation proceeds in the usual way: the  square of collinear matrix elements forms the PDFs~ in~(\ref{eq:dsigsq2}). 
	
The considerations are different when we start to investigate the NLP corrections. The ultimate reason for this is the presence of the time-ordered product operators, $J^{Tn}$, which can be used to provide power suppression as mentioned in Sec.~\ref{sec:SCET}. Insertions of the subleading-power Lagrangian terms introduce multiple threshold-collinear fields into the problem with an integral over the position of the insertion. The soft fields from the Lagrangian insertions are evaluated at the multipole expanded position.  This means that there is an extra convolution between the soft and collinear sectors which makes the threshold-collinear loops non-vanishing. We make these statements more concrete through the following example. 

We consider a term from the subleading-power Lagrangian 
	\begin{eqnarray}\label{eq:L2xi}
	\mathcal{L}^{(2)}_{2\xi}(z) =\frac{1}{2}\, \bar{\chi}_c(z)\,
	z^{\mu}_{\perp}\,z^{\nu}_{\perp}\,\Big[i\partial_{\nu}\,in_-\partial\,
	\mathcal{B}^+_{\mu}(z_-) \Big]\frac{\slashed{n}_{+}}{2}\,\chi_c(z) .
	\end{eqnarray}
The $\chi_c$ fields are the decoupled fields with the superscript $(0)$ dropped
and 
	\begin{eqnarray}\label{eq:softBB}
	\mathcal{B}_{\pm}^{\mu} &=& Y_{\pm}^{\dagger}
	\left[ i\,D^{\mu}_s\,Y_{\pm}\right] \,,
	\label{eq:softq} 
	\end{eqnarray}  
is the gauge invariant soft gauge field. We stress that the 
decoupling transformation has already been performed and  the soft-collinear interactions persist, as is clear from the Lagrangian term. 
The NLP analogue of equation (\ref{eq:2.2})
in this example is the following
	\begin{eqnarray}\label{1.19}
	\langle X|\bar{\psi} \gamma^\rho \psi(0)|A(p_A)B(p_B)\rangle
	&=&
	\int dt\, d\bar{t}\,  C^{A0,A0}(t, \bar{t}\,)\,
	\langle X^{{\rm PDF}}_{\bar{c}}|\bar{\chi}_{\bar{c},\alpha a}
	(\bar{t} n_-)|B(p_B)\rangle  
	\gamma_{\perp, \alpha \gamma}^\rho \nonumber\\ 
	&& \hspace*{-2.5cm} \times \,i\int d^4z \,\langle X^{{\rm PDF}}_{{c}}|
	\frac{1}{2}z_{\perp}^{\nu} z_{\perp}^{\mu} (in_-\partial_z)^2
	\,\mathbf{T}\left[ \chi_{c,\gamma f}\left(tn_+\right)
	\bar{\chi}_{c}\left(z\right)\textbf{T}^A \frac{\slashed n_+}{2} 
	\chi_{c}\left(z\right)\right]|A(p_A)\rangle \nonumber 
	\\ 
	&& \hspace*{-2.5cm}
	\times\,\langle X_s|   {\mathbf{ T}}
	\left( \left[ Y_{-}^\dagger(0) Y_{+}(0) \right]_{af}
	\frac{i\partial_{\perp}^{\mu}}{in_-\partial}
	\mathcal{B}^{+A}_{{\perp}\nu}\left(z_{-}\right)
	\right)|0 \rangle\,.\quad
	\end{eqnarray}
 In this equation, we see explicitly the extra convolution between the soft and collinear matrix elements, which motivates the non-vanishing  threshold-collinear loops and which induces the new threshold-collinear scale in the problem by injecting soft momentum into the collinear matrix element.
   
 The multiple threshold-collinear fields which are now present in (\ref{1.19}) due to the insertion of the subleading-power Lagrangian term,  cannot be radiated into the final state as in the threshold set up there is not enough energy available. On the contrary, the $c$-PDF modes, with $(Q,\Lambda^2/Q,\Lambda)$ scaling, {\it{can}} be radiated into the final state, just as at LP. Therefore, we define the {\it{NLP collinear function}} through the following 
 operator matching equation 
\begin{eqnarray}\label{m1}
i \int d^4z \,\mathbf{T}\Big[ 
\chi_{c,\gamma f}\left(tn_+\right)
\,\mathcal{L}^{(2)}(z) 
\Big] &&\nonumber \\&& \hspace{-2cm} 
= 2\pi \sum_i \int du 
\int dz_-
{	\tilde{J}_{i;\gamma\beta,\mu,fbd}
	\left(t,u;z_- \right)}\, 
\chi^{{\rm PDF}}_{c,\beta b}(un_+)
\,\textcolor{red}{\mathfrak{s}_{i;\mu,d}(z_-)}.
\end{eqnarray}
The indices $\mu$ and $d$ are collective Lorentz and colour
indices for each independent soft structure from  the set
 \begin{eqnarray}\label{s1}
 \textcolor{red}{\mathfrak{s}_{i}(z_-)} \in \left\{
 \textcolor{red}{\frac{i\partial_{\perp}^{\mu}}{in_-\partial}
 	\mathcal{B}^+_{\mu_\perp}(z_-)} \,,\,
 \textcolor{red}{
 	\frac{i\partial_{[\mu_{\perp}}}{in_-\partial}
 	\mathcal{B}^+_{\nu_\perp]}(z_-)} \,,\,
 \textcolor{red}{
 	\frac{1}{(in_-\partial)}
 	\big[  \mathcal{B}^+_{\mu_\perp}(z_-)
 	,\mathcal{B}^+_{\nu_\perp}(z_-)\,\big]}, .\,.\,.
 \right\}.
 \end{eqnarray} 
This operator equation defines formally the concept of  a
``radiative jet amplitude''~\cite{DelDuca:1990gz,Bonocore:2015esa,Bonocore:2016awd}.
The collinear function is formally a matching coefficient 
which can be computed perturbatively by considering 
partonic matrix elements of the above equation since  $Q\lambda=Q\sqrt{(1-z)} \gg  \Lambda$. 

\begin{figure}[t]
	\begin{centering}
		\includegraphics[width=0.44\textwidth]{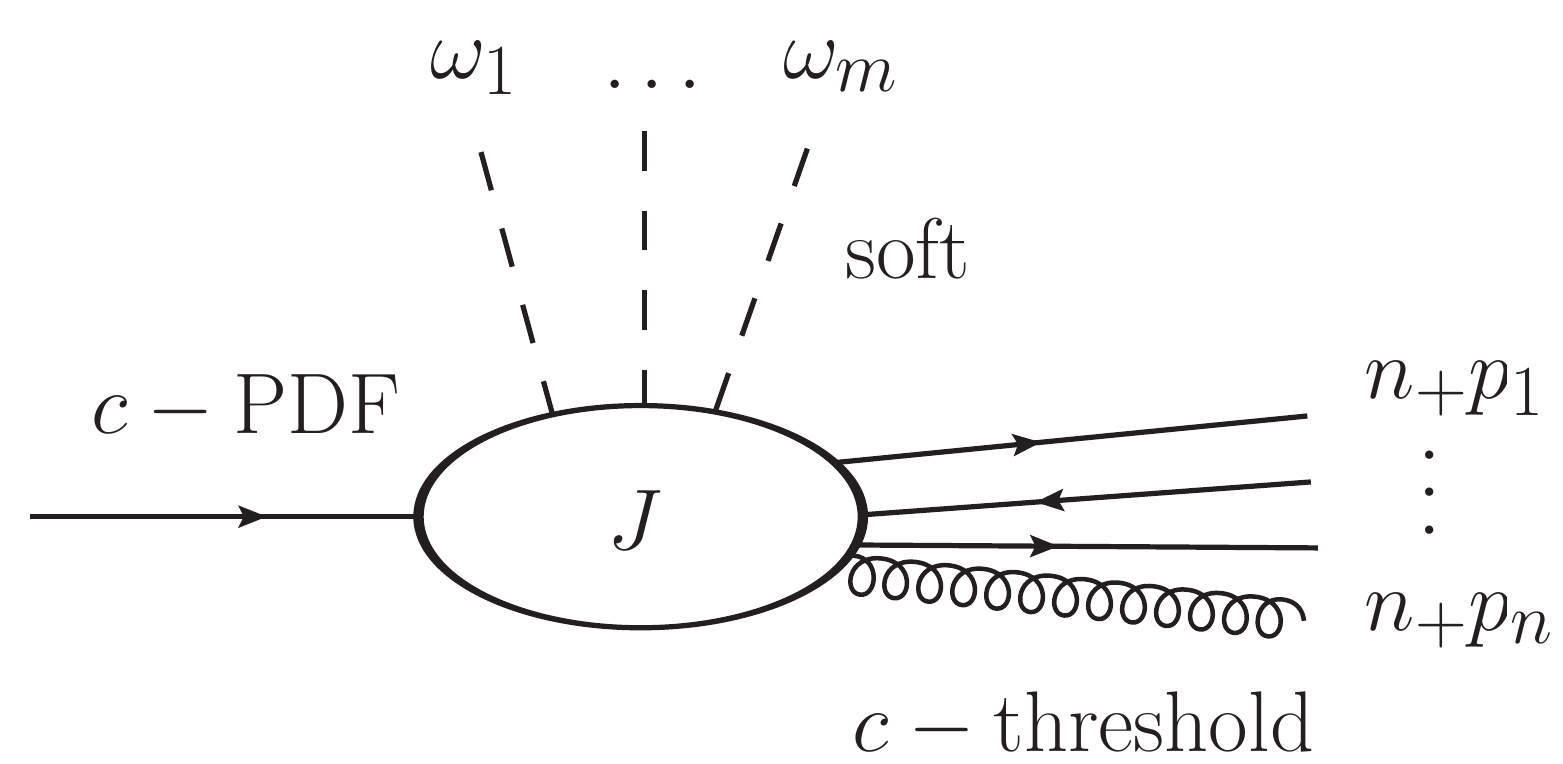}
		\par\end{centering}
	\caption{\label{fig:img4}   		
		A  pictorial, momentum-space, representation 
		of a general matching equation.  }
\end{figure}

We note that a more general version of this 
matching equation exists at general subleading powers.
This simply corresponds to including further insertions
of subleading-power Lagrangian terms on the left-hand 
side of the above equation. Moreover, even at NLP, instead
of an insertion of one term with power suppression of $\mathcal{O}(\lambda^2)$, there can appear 
two insertions of   $\mathcal{O}(\lambda^1)$ Lagrangian terms. In this case the
collinear and soft functions depend on $z_{1-}$ and $z_{2-}$, and appropriate integrals have to 
be added. A pictorial representation of a general momentum-space collinear function can be found in Fig.~\ref{fig:img4}. 
Since the general structure is a lengthy equation, and for us it is sufficient to consider (\ref{m1}),  we do not present the full form here but rather for more details we refer the reader to Sec. 2.3 of \cite{Beneke:2019oqx}.
\subsection{Generalized soft functions}
As we have seen in the previous section, explicit 
soft gauge fields from subleading-power Lagrangian terms
 play a passive role in the matching equation. The independent 
 collinear functions are defined with respect to the independent 
 soft structures. 
 Hence, at subleading powers the explicit soft gauge 
 fields are absorbed into the definition of soft functions, 
 extending the LP result in  (\ref{eq:LPsoftfn}), namely, we have 
\begin{eqnarray}
&&S_{i
}(\Omega,\omega) =
\int\frac{dx^0}{4\pi}e^{i x^0\Omega/2}\int \frac{dz_-}{2\pi}\,
e^{- i\omega  z_-}
\frac{1}{N_c}\mbox{Tr}
\langle 0 |\mathbf{\bar{T}}\left[Y_{+}^{\dagger}(x^0)Y_{-}(x^0)\right]
\mathbf{T}\left[Y_{-}^{\dagger}(0)Y_{+}(0) {\mathfrak{s}_{i}(z_-)}\right]|0\rangle.\quad\quad 
\end{eqnarray}
These soft functions, unlike their LP counterparts, exhibit divergences at their lowest non-vanishing order. 
This leads to an interesting renormalization group structure 
and mixing with new soft functions \cite{Beneke:2018gvs,Beneke:2019mua,Moult:2019uhz,Moult:2018jjd}.
However, we do not explore their rich structure further here. 
 \subsection{The factorization formula at NLP}
 Having discussed the new objects which appear 
 in the schematic factorization formula beyond LP 
 given in (\ref{eq:factsketch}), we are in position
 to derive the precise form of factorization at NLP.
  We do not do this in detail here, as this derivation 
 is rather technical and was carefully presented in
 Sec. 3 of \cite{Beneke:2019oqx}. We simply state that
 the NLP cross section is split into two contributions,
  the {\it{kinematic}}, where power corrections come from the expansion of phase space, 
  and {\it{dynamical}}, which in turn are due to explicit insertions of subleading-power 
  Lagrangian terms. We choose to define 
  \begin{eqnarray} 
  \Delta(z) = \frac{1}{(1-\epsilon)} \frac{\hat{\sigma}(z)}{z}\,, 
  \end{eqnarray}
 which makes the comparison with existing literature easier. 
  The $\epsilon$ still appears as the factorization formula 
 we derive is for $d$-dimensional regularized quantities.   

One of the main results included in this contribution is the
following all-order formulation of NLP factorization  
\begin{eqnarray}
\label{eq:3.24}
\Delta^{dyn }_{{\rm{NLP}} }(z)&=&- \frac{2}{(1-\epsilon)} \,  
Q \left[ \left(\frac{\slashed{n}_-}{4}
\right) {\gamma}_{\perp\rho}  \left(\frac{\slashed{n}_+}{4}
\right) \gamma^{\rho}_{\perp} \right]_{\beta\gamma}
\nonumber  \\ 
&&  \hspace{0cm} \times 
\, \int d(n_+p)\,C^{\,{A0,A0}\,}(n_+p, x_bn_-{p}_B  ) \,
C^{*A0A0}\left(\,x_an_+p_A,\,x_b{n_-p_B}\right)
\nonumber \\ 
&&  \hspace{0cm} \times \,\sum^5_{i=1}
\,\int \left\{d\omega_j\right\}
{J}_{i,\gamma\beta}\left(n_+p,x_a n_+p_A; 
\left\{\omega_j\right\} \right) \, 
{S}_{i}(\Omega; \left\{\omega_j\right\} ) 
+\rm{h.c.}\,,\quad
\end{eqnarray}
where here $\Omega=Q(1-z)$. 
The collinear functions are written in momentum space.
 We do not list 
all of the soft functions present in the
above formula here, these can be found in 
Eqs.~(3.34) - (3.39) of \cite{Beneke:2019oqx}. 
The spin trace present in LP result cannot 
yet be performed at this stage as there exist
soft and collinear structures which share Dirac 
indices and must be contracted prior to performing the spin trace. 

As pointed out above, the soft structures begin at the $\alpha_s$ order
because of the explicit insertion of the soft fields. Interestingly,
only one allowed soft structure contains exactly one soft gauge field.
We will focus on this contribution in what follows. 
The soft function is given by 
\begin{eqnarray}
{S}_{1}(\Omega; \omega ) &=& 
\int \frac{dx^0}{4\pi} \,  e^{i \Omega\, x^0/2}  
\int \frac{dz_{j-}}{2\pi}  \, e^{-i\omega {z_{-}}}
\nonumber\\\label{eq:3.23} &&\times
\frac{1}{N_c}\, {\rm{Tr}} \langle 0| \bar{\mathbf{ T}} \left[ Y_{+}^\dagger(x^0)
Y_{-}(x^0)  \right] {\mathbf{ T}}\left(
\left[ Y_{-}^\dagger(0) Y_{+}(0) \right]
\frac{i\partial_{\perp}^{\nu}}{in_-\partial}
\mathcal{B}^{+}_{\nu_{\perp}}\left(z_{-}\right)
\right)|0 \rangle\,, \quad
\end{eqnarray}
with the one-loop result (presented in expanded form in \cite{Beneke:2018gvs})
\begin{eqnarray}\label{eq:softfunc}
S^{(1)}_{1}\left(\Omega,\omega\right) = \frac{\alpha_sC_F}{2\pi}
\frac{\mu^{2\epsilon}e^{\epsilon\gamma_E}}{\Gamma[1-\epsilon]}
\,\frac{1}{\omega^{1+\epsilon}}\frac{1}{(\Omega-\omega)^{\epsilon}}
\,\theta(\omega)\theta(\Omega-\omega) \,\,.\quad
\end{eqnarray}
This contribution is of interest to us, because 
the fact that it is the only soft structure which 
begins at $\mathcal{O}(\alpha_s)$ means that  
the next-to-leading order (NLO) contribution is 
determined by it, and tree-level values for the
hard function, $H(\hat{s})=|C^{A0,A0}(x_an_+p_A,x_bn_-p_B)|^2$,
and collinear function, $J_1$. Moreover, $J_1\otimes S_1$ is also 
the only contribution to the collinear sector at next-to-next-leading
order (NNLO), as the rest of the soft functions begin already at $\mathcal{O}(\lambda^2)$ and addition of a collinear loop yields N$^3$LO terms. This is useful as we would like to study the new 
NLP collinear functions at one-loop order, and using this we can 
check the validity of our results to NNLO. 
 We will discuss $J_1$ in more 
details in the next section. 

We have neglected the kinematic corrections, 
$\Delta^{kin}$ since its structure is simpler than one of~$\Delta^{dyn}$.
However, it has been included in \cite{Beneke:2019oqx}.
\section{Collinear function at one-loop order and fixed-order check}
The result for $J_1$ collinear function is another main result 
of this contribution and \cite{Beneke:2019oqx}. 
The calculation is a matching computation the operator equation in (\ref{m1}).
It is rather involved and can be found in detail in Sec. 4 of \cite{Beneke:2019oqx}. 
Here we give the results and discuss the implications. 
The $J_1$ collinear function (with index structure restored for completeness) 
is given by 
\begin{eqnarray}
{J}^{K(0)}_{1;\gamma\beta,fq}(n_+q,n_+p;\omega)
&=&   \mathbf{T}^{K}_{fq}  \delta_{\beta\gamma } 
\left(-\frac{1}{n_+p} \delta(n_+q -n_+p)  
+2 \,\frac{\partial}{\partial n_+q}
\delta(n_+q -n_+p) \right),\qquad\,\,
\label{eq:J1fn}
\end{eqnarray}
at tree level, and 
\begin{eqnarray}
\label{J1}
&& {J}^{K\,(1)}_{1;\gamma\beta, fq}
\left(n_+q, n_+p; \,\omega \right) = 
\frac{\alpha_s}{4 \pi} 
\delta_{\gamma\beta}\mathbf{T}_{fq}^{K} 
\, \frac{1}{(n_+p) } \left(\frac{n_+p\,\omega}
{\mu ^2}\right)^{-\epsilon }
\frac{e^{\epsilon\,\gamma_E}\,\Gamma[1+\epsilon ]
	\Gamma [1-\epsilon]^2}
{(-1+\epsilon)(1+\epsilon) \Gamma [2-2 \epsilon ]}
\nonumber\\ 
&& \hspace{1.5cm}\times  
\left( C_F\left(-\frac{4}{\epsilon}+3 
+8\epsilon+\epsilon^2  \right)
-  C_A \left(-5+{8}{\epsilon}
+\epsilon ^2 \right)\right) \delta(n_+q -n_+p)\qquad
\\
&& \hspace{1cm} = \label{J1expanded}
\frac{ \alpha_s}{4 \pi} \,
\frac{1}{(n_+p)}  \, \delta_{\gamma\beta}
\mathbf{T}_{fq}^{K}
\bigg( C_F\left(\frac{4}{\epsilon }+5-4 \ln
\left(\frac{n_+p\,\omega}{\mu^2}\right) \right) 
-  5\, C_A \bigg)\,\delta(n_+q -n_+p)  
 +\,\mathcal{O}(\epsilon) \,,\qquad
\end{eqnarray}
at one-loop order. We have also expanded the result 
in $\epsilon$ for discussion. 
The NNLO collinear  contribution is 
given by 
	\begin{eqnarray}\label{eq:3.33}
\Delta^{dyn\,(2)}_{{\rm{NLP-coll}}}(z)&=& 4  Q \, 
H^{(0)}\big(Q^2\big) \int d\omega
\,  {J}^{\,(1)}_{1}
\left(x_an_+p_A; \omega \right) \, 
{S}^{\,(1)}_{1}(\Omega; \omega ) \,,
\end{eqnarray}
where the delta function from the collinear function has been used already.
Inserting $J_1$ from (\ref{J1}) and $S_1$ from (\ref{eq:softfunc}) 
and performing the last integral gives 
\begin{eqnarray}\label{sig9}
\Delta^{dyn\,(2)}_{{\rm{NLP-coll}}}(z)&=& 
\frac{  \alpha_s^2 }{(4\pi)^2} 
\Bigg(
{C_A}C_F
\left( \frac{20}{\epsilon }-(60 \ln (1-z)-8)
+\mathcal{O}(\epsilon)  \right) \\
&&\hspace{-2cm} + 
C_F^2 \,\bigg(-\frac{16}{\epsilon ^2}
+\frac{48 \ln (1-z)-20}{\epsilon } \nonumber  
+ 
\left(-72 \ln ^2(1-z)+60 \ln (1-z)+8 \pi ^2-24\right) 
+\mathcal{O}(\epsilon)  \bigg) \,\Bigg)\,.
\end{eqnarray} 
where we set $\mu=Q$. We note that leading logarithms, $\sim\alpha^2_s\ln^3(1-z)$ do not appear which is an indication that the definition used for collinear function is consistent \cite{Beneke:2018gvs}. 
 The $C_F^2$ term in~(\ref{sig9}) 
 is in agreement with the corresponding abelian contribution considered in Eq.~(4.22) 
 of \cite{Bonocore:2015esa} and
 Eqs.~(13), (14) of \cite{Bonocore:2014wua} in the diagrammatic and expansion-by-regions methods respectively.
\section{Ill-defined convolution}
The last remark we wish to make,  is with 
relation to the $\epsilon$ expanded result. To obtain Eq.~(\ref{sig9})
we have used $d$-dimensional quantities $J_1$ and $S_1$, performed the
$\omega$ integral first and then expanded in $\epsilon$. If this order is
reversed, that is, if we convolve renormalized objects, ill-defined terms such as $\int d\omega \,\delta(\omega)\, \ln(\omega)$ appear. 
This is an issue for extending NLP resummation beyond leading logarithmic 
order and is a open interesting conceptual challenge in the community, see for example \cite{Moult:2019uhz}.
With the results presented here we see the issue explicitly.  
\section{Summary}
In this contribution we have outlined the formalism which 
can be used to describe general processes at subleading powers 
in $\lambda$ expansion. We have then used this formalism, focusing
on the case of DY to motivate interesting structure of factorization
at NLP, where new objects, the {\it{NLP collinear functions}} appear. 
We have also presented the new results for one-loop collinear function,
using which the ill-defined convolution issue can be clearly seen. 
This contribution is meant as an overview of the formalism, with 
new interesting features mentioned but not derived step by step. 
A fuller, more detailed discussion is presented in \cite{Beneke:2019oqx}.
\section*{Acknowledgments}
I would like to thank M. Beneke, A. Broggio, R. Szafron, and L. Vernazza for careful reading of the manuscript and useful suggestions. 
This work has been supported by the Bundesministerium f\"ur 
Bildung and Forschung (BMBF) grant no. 05H18WOCA1.

\bibliographystyle{JHEP}
\bibliography{NLP}

\providecommand{\href}[2]{#2}\begingroup\raggedright\begin{thebibliography}{10}

\bibitem{Beneke:2019oqx}
M.~Beneke, A.~Broggio, S.~Jaskiewicz and L.~Vernazza, \emph{{Threshold
  factorization of the Drell-Yan process at next-to-leading power}},
  \href{https://arxiv.org/abs/1912.01585}{{\ttfamily 1912.01585}}.

\bibitem{Beneke:2018gvs}
M.~Beneke, A.~Broggio, M.~Garny, S.~Jaskiewicz, R.~Szafron, L.~Vernazza et~al.,
  \emph{{Leading-logarithmic threshold resummation of the Drell-Yan process at
  next-to-leading power}},
  \href{https://doi.org/10.1007/JHEP03(2019)043}{\emph{JHEP} {\bfseries 03}
  (2019) 043}, [\href{https://arxiv.org/abs/1809.10631}{{\ttfamily
  1809.10631}}].

\bibitem{Beneke:2004in}
M.~Beneke, F.~Campanario, T.~Mannel and B.~D. Pecjak, \emph{{Power corrections
  to $\bar{B} \to X_u \ell \bar{\nu} \,(X_s \gamma)$ decay spectra in the
  `shape-function' region}},
  \href{https://doi.org/10.1088/1126-6708/2005/06/071}{\emph{JHEP} {\bfseries
  06} (2005) 071}, [\href{https://arxiv.org/abs/hep-ph/0411395}{{\ttfamily
  hep-ph/0411395}}].

\bibitem{Beneke:2019mua}
M.~Beneke, M.~Garny, S.~Jaskiewicz, R.~Szafron, L.~Vernazza and J.~Wang,
  \emph{{Leading-logarithmic threshold resummation of Higgs production in gluon
  fusion at next-to-leading power}},
  \href{https://arxiv.org/abs/1910.12685}{{\ttfamily 1910.12685}}.

\bibitem{Beneke:2017vpq}
M.~Beneke, C.~Bobeth and R.~Szafron, \emph{{Enhanced electromagnetic correction
  to the rare $B$-meson decay $B_{s,d} \to \mu^+ \mu^-$}},
  \href{https://doi.org/10.1103/PhysRevLett.120.011801}{\emph{Phys. Rev. Lett.}
  {\bfseries 120} (2018) 011801},
  [\href{https://arxiv.org/abs/1708.09152}{{\ttfamily 1708.09152}}].

\bibitem{Beneke:2019slt}
M.~Beneke, C.~Bobeth and R.~Szafron, \emph{{Power-enhanced leading-logarithmic
  QED corrections to $B_q \to \mu^+\mu^-$}},
  \href{https://doi.org/10.1007/JHEP10(2019)232}{\emph{JHEP} {\bfseries 10}
  (2019) 232}, [\href{https://arxiv.org/abs/1908.07011}{{\ttfamily
  1908.07011}}].

\bibitem{Alte:2018nbn}
S.~Alte, M.~Koenig and M.~Neubert, \emph{{Effective Field Theory after a
  New-Physics Discovery}},
  \href{https://doi.org/10.1007/JHEP08(2018)095}{\emph{JHEP} {\bfseries 08}
  (2018) 095}, [\href{https://arxiv.org/abs/1806.01278}{{\ttfamily
  1806.01278}}].

\bibitem{Moult:2019uhz}
I.~Moult, I.~W. Stewart, G.~Vita and H.~X. Zhu, \emph{{The Soft Quark
  Sudakov}},  \href{https://arxiv.org/abs/1910.14038}{{\ttfamily 1910.14038}}.

\bibitem{Beneke:2002ph}
M.~Beneke, A.~P. Chapovsky, M.~Diehl and T.~Feldmann, \emph{{Soft collinear
  effective theory and heavy to light currents beyond leading power}},
  \href{https://doi.org/10.1016/S0550-3213(02)00687-9}{\emph{Nucl. Phys.}
  {\bfseries B643} (2002) 431--476},
  [\href{https://arxiv.org/abs/hep-ph/0206152}{{\ttfamily hep-ph/0206152}}].

\bibitem{Beneke:2002ni}
M.~Beneke and T.~Feldmann, \emph{{Multipole expanded soft collinear effective
  theory with non-abelian gauge symmetry}},
  \href{https://doi.org/10.1016/S0370-2693(02)03204-5}{\emph{Phys. Lett.}
  {\bfseries B553} (2003) 267--276},
  [\href{https://arxiv.org/abs/hep-ph/0211358}{{\ttfamily hep-ph/0211358}}].

\bibitem{Bauer:2000yr}
C.~W. Bauer, S.~Fleming, D.~Pirjol and I.~W. Stewart, \emph{{An Effective field
  theory for collinear and soft gluons: Heavy to light decays}},
  \href{https://doi.org/10.1103/PhysRevD.63.114020}{\emph{Phys. Rev.}
  {\bfseries D63} (2001) 114020},
  [\href{https://arxiv.org/abs/hep-ph/0011336}{{\ttfamily hep-ph/0011336}}].

\bibitem{Bauer:2001yt}
C.~W. Bauer, D.~Pirjol and I.~W. Stewart, \emph{{Soft collinear factorization
  in effective field theory}},
  \href{https://doi.org/10.1103/PhysRevD.65.054022}{\emph{Phys. Rev.}
  {\bfseries D65} (2002) 054022},
  [\href{https://arxiv.org/abs/hep-ph/0109045}{{\ttfamily hep-ph/0109045}}].

\bibitem{Bauer:2001ct}
C.~W. Bauer and I.~W. Stewart, \emph{{Invariant operators in collinear
  effective theory}},
  \href{https://doi.org/10.1016/S0370-2693(01)00902-9}{\emph{Phys. Lett.}
  {\bfseries B516} (2001) 134--142},
  [\href{https://arxiv.org/abs/hep-ph/0107001}{{\ttfamily hep-ph/0107001}}].

\bibitem{Beneke:2017ztn}
M.~Beneke, M.~Garny, R.~Szafron and J.~Wang, \emph{{Anomalous dimension of
  subleading-power N-jet operators}},
  \href{https://doi.org/10.1007/JHEP03(2018)001}{\emph{JHEP} {\bfseries 03}
  (2018) 001}, [\href{https://arxiv.org/abs/1712.04416}{{\ttfamily
  1712.04416}}].

\bibitem{Beneke:2018rbh}
M.~Beneke, M.~Garny, R.~Szafron and J.~Wang, \emph{{Anomalous dimension of
  subleading-power $N$-jet operators. Part II}},
  \href{https://doi.org/10.1007/JHEP11(2018)112}{\emph{JHEP} {\bfseries 11}
  (2018) 112}, [\href{https://arxiv.org/abs/1808.04742}{{\ttfamily
  1808.04742}}].

\bibitem{Beneke:2019kgv}
M.~Beneke, M.~Garny, R.~Szafron and J.~Wang, \emph{{Violation of the
  Kluberg-Stern-Zuber theorem in SCET}},
  \href{https://doi.org/10.1007/JHEP09(2019)101}{\emph{JHEP} {\bfseries 09}
  (2019) 101}, [\href{https://arxiv.org/abs/1907.05463}{{\ttfamily
  1907.05463}}].

\bibitem{Feige:2017zci}
I.~Feige, D.~W. Kolodrubetz, I.~Moult and I.~W. Stewart, \emph{{A Complete
  Basis of Helicity Operators for Subleading Factorization}},
  \href{https://doi.org/10.1007/JHEP11(2017)142}{\emph{JHEP} {\bfseries 11}
  (2017) 142}, [\href{https://arxiv.org/abs/1703.03411}{{\ttfamily
  1703.03411}}].

\bibitem{Chang:2017atu}
C.-H. Chang, I.~W. Stewart and G.~Vita, \emph{{A Subleading Power Operator
  Basis for the Scalar Quark Current}},
  \href{https://doi.org/10.1007/JHEP04(2018)041}{\emph{JHEP} {\bfseries 04}
  (2018) 041}, [\href{https://arxiv.org/abs/1712.04343}{{\ttfamily
  1712.04343}}].

\bibitem{Ebert:2018lzn}
M.~A. Ebert, I.~Moult, I.~W. Stewart, F.~J. Tackmann, G.~Vita and H.~X. Zhu,
  \emph{{Power Corrections for N-Jettiness Subtractions at ${\cal
  O}(\alpha_s)$}}, \href{https://doi.org/10.1007/JHEP12(2018)084}{\emph{JHEP}
  {\bfseries 12} (2018) 084},
  [\href{https://arxiv.org/abs/1807.10764}{{\ttfamily 1807.10764}}].

\bibitem{Moult:2017rpl}
I.~Moult, I.~W. Stewart and G.~Vita, \emph{{A subleading operator basis and
  matching for gg $\to$ H}},
  \href{https://doi.org/10.1007/JHEP07(2017)067}{\emph{JHEP} {\bfseries 07}
  (2017) 067}, [\href{https://arxiv.org/abs/1703.03408}{{\ttfamily
  1703.03408}}].

\bibitem{Moult:2018jjd}
I.~Moult, I.~W. Stewart, G.~Vita and H.~X. Zhu, \emph{{First Subleading Power
  Resummation for Event Shapes}},
  \href{https://doi.org/10.1007/JHEP08(2018)013}{\emph{JHEP} {\bfseries 08}
  (2018) 013}, [\href{https://arxiv.org/abs/1804.04665}{{\ttfamily
  1804.04665}}].

\bibitem{Becher:2007ty}
T.~Becher, M.~Neubert and G.~Xu, \emph{{Dynamical Threshold Enhancement and
  Resummation in Drell- Yan Production}},
  \href{https://doi.org/10.1088/1126-6708/2008/07/030}{\emph{JHEP} {\bfseries
  07} (2008) 030}, [\href{https://arxiv.org/abs/0710.0680}{{\ttfamily
  0710.0680}}].

\bibitem{Korchemsky:1993uz}
G.~P. Korchemsky and G.~Marchesini, \emph{{Resummation of large infrared
  corrections using Wilson loops}},
  \href{https://doi.org/10.1016/0370-2693(93)90015-A}{\emph{Phys. Lett.}
  {\bfseries B313} (1993) 433--440}.

\bibitem{Moch:2005ky}
S.~Moch and A.~Vogt, \emph{{Higher-order soft corrections to lepton pair and
  Higgs boson production}},
  \href{https://doi.org/10.1016/j.physletb.2005.09.061}{\emph{Phys. Lett.}
  {\bfseries B631} (2005) 48--57},
  [\href{https://arxiv.org/abs/hep-ph/0508265}{{\ttfamily hep-ph/0508265}}].

\bibitem{DelDuca:1990gz}
V.~Del~Duca, \emph{{High-energy Bremsstrahlung Theorems for Soft Photons}},
  \href{https://doi.org/10.1016/0550-3213(90)90392-Q}{\emph{Nucl. Phys.}
  {\bfseries B345} (1990) 369--388}.

\bibitem{Bonocore:2015esa}
D.~Bonocore, E.~Laenen, L.~Magnea, S.~Melville, L.~Vernazza and C.~D. White,
  \emph{{A factorization approach to next-to-leading-power threshold
  logarithms}}, \href{https://doi.org/10.1007/JHEP06(2015)008}{\emph{JHEP}
  {\bfseries 06} (2015) 008},
  [\href{https://arxiv.org/abs/1503.05156}{{\ttfamily 1503.05156}}].

\bibitem{Bonocore:2016awd}
D.~Bonocore, E.~Laenen, L.~Magnea, L.~Vernazza and C.~D. White,
  \emph{{Non-abelian factorisation for next-to-leading-power threshold
  logarithms}}, \href{https://doi.org/10.1007/JHEP12(2016)121}{\emph{JHEP}
  {\bfseries 12} (2016) 121},
  [\href{https://arxiv.org/abs/1610.06842}{{\ttfamily 1610.06842}}].

\bibitem{Bonocore:2014wua}
D.~Bonocore, E.~Laenen, L.~Magnea, L.~Vernazza and C.~D. White, \emph{{The
  method of regions and next-to-soft corrections in Drell-Yan production}},
  \href{https://doi.org/10.1016/j.physletb.2015.02.008}{\emph{Phys. Lett.}
  {\bfseries B742} (2015) 375--382},
  [\href{https://arxiv.org/abs/1410.6406}{{\ttfamily 1410.6406}}].

\end{thebibliography}\endgroup

\end{document}